\documentclass[prb,twocolumn]{revtex4-1}

\usepackage{amsmath}
\usepackage{amsfonts}
\usepackage{graphicx}

\begin{document}

\title{Interactive molecular dynamics}

\author{Daniel V. Schroeder}
\email{dschroeder@weber.edu}
\affiliation{Physics Department, Weber State University, Ogden, UT 84408-2508}

\begin{abstract}

Physics students now have access to interactive molecular dynamics simulations that can model and animate the motions of hundreds of particles, such as noble gas atoms, that attract each other weakly at short distances but repel strongly when pressed together.  Using these simulations, students can develop an understanding of forces and motions at the molecular scale, nonideal fluids, phases of matter, thermal equilibrium, nonequilibrium states, the Boltzmann distribution, the arrow of time, and much more.  This article summarizes the basic features and capabilities of such a simulation, presents a variety of student exercises using it at the introductory and intermediate levels, and describes some enhancements that can further extend its uses.  A working simulation code, in HTML5 and JavaScript for running within any modern Web browser, is provided as an online supplement.\textsl{Copyright \copyright 2014, American Association of Physics Teachers.} Published in Am.\ J.\ Phys.\ \textbf{83} (3), 210--218 (2015), \url{<http://dx.doi.org/10.1119/1.4901185>}.

\end{abstract}

\maketitle

\section{Introduction}

The atomic theory of matter is a pillar of modern science. Richard Feynman said it best:\cite{FeynmanLectures, FeynmanVideo, VMDL}
\begin{quote}
If, in some cataclysm, all of scientific knowledge were to be destroyed, and only one sentence passed on to the next generations of creatures, what statement would contain the most information in the fewest words? I believe it is the \textit{atomic hypothesis} (or the atomic \textit{fact}, or whatever you wish to call it) that \textit{all things are made of atoms---little particles that move around in perpetual motion, attracting each other when they are a little distance apart, but repelling upon being squeezed into one another}. In that one sentence, you will see, there is an \textit{enormous} amount of information about the world, if just a little imagination and thinking are applied.
\end{quote}
Experienced physicists can readily apply ``just a little imagination and thinking'' to this view of matter, to arrive at a qualitative, microscopic understanding of gas pressure, phase changes, thermal equilibrium, nonequilibrium states, irreversible behavior, and specific phenomena such as friction, thermal expansion, surface tension, crystal dislocations, and Brownian motion. With further analysis and calculation, physicists can also quantify much of this understanding in terms of the laws of thermodynamics, Boltzmann statistics, kinetic theory, and so on.

Non-experts, however, do not automatically imagine and think about these phenomena correctly.  For example, many students find it difficult to picture atoms in \textit{perpetual} motion, colliding elastically with no loss of energy over time---even though this picture is essential to understanding how a gas can exert a steady pressure.\cite{NovickNussbaum, MakingSense, Arons}
Physics students gradually learn quantitative approaches to thermodynamics and statistical mechanics, but their qualitative understanding can lag behind their symbol-pushing skills.  Moreover, at the undergraduate level, the thermal physics curriculum tends to be restricted to a disappointingly narrow range of systems:  
\begin{itemize}
\item ``ideal'' gases in which the particles don't attract or repel each other at all; and/or
\item systems in equilibrium (so that the fundamental assumption of statistical mechanics applies); and/or
\item very large systems (in the ``thermodynamic limit''), for which fluctuations and surface effects are negligible.
\end{itemize}
Of course, these restrictions are usually necessary in order to perform accurate pencil-and-paper calculations.

To study systems that are free of these restrictions, researchers often use molecular dynamics simulations.\cite{AllenAndTildesley, Haile, Rapaport}  In their basic form, these simulations integrate Newton's second law numerically to determine the motions of a moderately large number of classical particles.  In this approach there is no difficulty with incorporating forces between the particles, or with studying nonequilibrium states.  Simulating very large systems is computationally expensive, so fluctuations and surface effects are necessarily apparent (for better or for worse) in practical simulations.

Molecular dynamics is gradually making its way into science education, from two directions.  At the elementary level, there are now simulations with animated graphics that are intended to give precollege students and chemistry students a qualitative understanding of the atomic view of matter, and especially of the differences between solids, liquids, and gases.\cite{AtomicMicroscope,AtomicMicroscopeReview,AtomsInMotion,mw,phet}  Meanwhile, at a more advanced level, a growing number of text materials\cite{GouldTobochnikChristian,Giordano,DVSJavaManual,Sander} for computational physics and statistical physics are presenting molecular dynamics template codes and encouraging students to modify them and use them to perform quantitative numerical ``experiments.''

The purpose of this article is to encourage more widespread use of molecular dynamics simulations in physics instruction at all levels, so that more students will understand the atomic view of matter and learn to apply it, both qualitatively and quantitatively, to a wider variety of phenomena.  There are especially many opportunities to incorporate molecular dynamics simulations into courses in introductory physics and thermal physics.  In these courses the students rarely have time to do their own coding, yet they are ready to appreciate the principles behind a molecular dynamics simulation and, with sufficiently flexible software, to use it to gain solid qualitative understanding and to conduct serious numerical experiments.  Software intended for this type of student use does exist,\cite{GouldTobochnikStat, OSPApplet, VMDL} but it is not widely used and in my opinion there is a need for a greater diversity of software options.  In an attempt to partially fill this need, I have written an interactive molecular dynamics simulation with animated graphics in HTML5/JavaScript, which is provided as an online supplement to this article.\cite{theSimulation}  Readers may wish to run this simulation while reading the rest of the article.  There is no need to view the source code of this simulation, but I have tried to make the code easy for beginning programmers to read, in the hope that curious students will look to see how it works, and in the hope that other instructors will modify and adapt it to fulfill a still wider variety of educational needs.

The next two sections summarize the Lennard-Jones model that is widely used in molecular dynamics simulations, and discuss some details of implementing the model on a computer.  These sections mostly review material that can readily be found elsewhere, but are included for completeness.  Section IV then briefly summarizes some of the qualitative behavior of the simulated Lennard-Jones system.  Section V lists several user interface features that facilitate extensive and open-ended exploration of the system's behavior.  A collection of 20 student exercises is presented in Section VI, and further enhancements to a basic molecular dynamics simulation are briefly described in Section~VII.

\section{The Lennard-Jones model}

\begin{figure}[t]
\centering
\includegraphics[width=7.5cm]{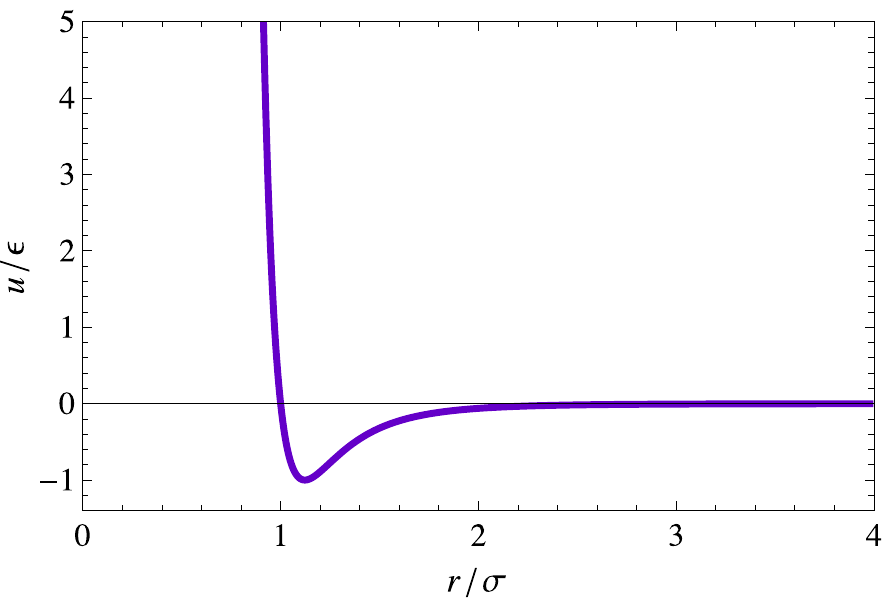}
\caption{The Lennard-Jones 6-12 potential energy function, Eq.~(\ref{LJ}).}
\label{LennardJonesPlot}
\end{figure}

In the simplest molecular dynamics simulations, the model consists of $N$ classical, spherically symmetrical particles interacting pair-wise. A commonly used form for the interaction\cite{AllenAndTildesley, Haile, Rapaport, GouldTobochnikChristian, Giordano} is the Lennard-Jones 6-12 potential energy function,
\begin{equation}
\label{LJ}
u(r) = 4\epsilon\biggl[\Bigl(\frac\sigma{r}\Bigr)^{12} - \Bigl(\frac\sigma{r}\Bigr)^6\biggr],
\end{equation}
where $r$ is the distance between the centers of the two particles and $\sigma$ and $\epsilon$ are constants.  This function is plotted in Fig.~\ref{LennardJonesPlot}, where we see that the potential reaches a minimum value of $-\epsilon$ at $r=2^{1/6}\,\sigma$ and rises very steeply for $r<\sigma$.  Thus we can think of $\sigma$ as a rough measure of the diameter of the particles.  The attractive $r^{-6}$ term in Eq.~(\ref{LJ}) represents the long-range behavior of the van der Waals force between uncharged, nonpolar molecules.\cite{london}  There is no good theoretical or experimental basis for the exact form of the repulsive $r^{-12}$ term, which is chosen for computational convenience.  In any case, this pair-wise potential is a reasonable semi-quantitative model for the interactions of noble gas atoms.\cite{maitland}

\begin{table}[b]
\centering
\caption{Approximate sizes of natural units of the Lennard-Jones system, when this system is used to model various noble gases. Values of $\sigma$ and $\epsilon$ are adapted from Ref.~\onlinecite{maitland}, p.~582. The $\epsilon$ values can vary by 10\% or more, depending on what experimental data are used to fit the Lennard-Jones function.}
\begin{ruledtabular}
\begin{tabular}{l c c c c c c}
& Mass & Length & Energy & Temp. & Velocity & Time \\
& $m$ & $\sigma$ & $\epsilon$ & $\epsilon/k_B$ & $\sqrt{\epsilon/m}$ & $\sigma\sqrt{m/\epsilon}$ \\
& (u) & (nm) & (eV) & (K) & (m/s) & (ps) \\
\hline
Helium & 4.0 & 0.264 & 0.00094 & 10.9 & 150 & 1.76 \\
Neon & 20.2 & 0.274 & 0.0035 & 41.2 & 130 & 2.10 \\
Argon & 39.9 & 0.335 & 0.0122 & 142 & 172 & 1.95 \\
Krypton & 83.8 & 0.358 & 0.0172 & 199 & 141 & 2.55 \\
Xenon & 131.3 & 0.380 & 0.0242 & 281 & 133 & 2.84 \\
\end{tabular}
\end{ruledtabular}
\label{bosons}
\end{table}

Computational physicists generally use (with no loss of generality) a natural system of units in which the Lennard-Jones parameters $\sigma$ and $\epsilon$ are both equal to~1, along with the particle mass~$m$ and Boltzmann's constant~$k_B$:
\begin{equation}
\sigma = \epsilon = m = k_B = 1.
\end{equation}
This article and the accompanying code also use this natural unit system, even though the intent is to reach students who may have never worked physics problems in non-SI units. Working in natural units and converting between unit systems are essential skills that all scientists and engineers must learn at some point.  Still, in some educational settings it will be appropriate to work in more familiar units.  Table~I lists some approximate values of Lennard-Jones parameters that can be used to model different noble gases, including the natural units of temperature ($\epsilon/k_B$), velocity ($\sqrt{\epsilon/m}$), and time ($\sigma\sqrt{m/\epsilon}$).

To keep the particles from drifting outward indefinitely, a simulation must either use periodic boundary conditions or provide an additional confining force.  While the former choice is preferable for many research studies, students find it easier to conceptualize molecules that bounce off of walls.\cite{WallsComment}  Using walls also offers some other advantages, as described below.

It would be most natural to model atoms moving in three-dimensional space, and we could then make quantitative comparisons to experimental data for real three-dimensional noble gases and their condensed phases.  However, a two-dimensional simulation can demonstrate most of the important physical principles equally well, and is preferable for many educational purposes.  Then the student can see all the simulated atoms at once on a graphical display, and interact with them in a natural way using a mouse or other pointing mechanism.  Both graphics and pointing interactions become awkward, though certainly not impossible, with a three-dimensional simulation.  A two-dimensional simulation also typically requires fewer particles, so it can run at a faster animation rate.

In summary, this article is primarily about two-dimensional dynamical Lennard-Jones simulations with fixed-wall boundary conditions.

\section{Computational details}

For those who are interested in understanding the computational algorithms of a molecular dynamics simulation, and perhaps coding their own simulations, abundant resources are available.\cite{AllenAndTildesley, Haile, Rapaport, GouldTobochnikChristian, Giordano}  The following remarks merely provide a brief overview and highlight a few important issues.

To integrate Newton's second law for the $N$-particle Lennard-Jones system, a common (and easy) approach is to use the velocity Verlet algorithm.  The program stores the positions ($\vec r_i$), velocities ($\vec v_i$), and accelerations ($\vec a_i$) of the $N$ particles, and repeatedly steps these variables forward by small time increments ($dt$) in the following sequence:
\begin{enumerate}
\item Use the current velocities and accelerations to update all of the positions to second-order accuracy: $\vec r_i \longleftarrow \vec r_i + \vec v_i\,dt + \frac12 \vec a_i(dt)^2$.
\item Use the current accelerations to update all of the velocities by half a time step: $\vec v_i \longleftarrow \vec v_i + \frac12 \vec a_i\,dt$.
\item Compute the new accelerations from the updated positions, using the Lennard-Jones force law (i.e., the gradient of Eq.~(\ref{LJ})).
\item Use these new accelerations to update the velocities by another half time step: $\vec v_i \longleftarrow \vec v_i + \frac12 \vec a_i\,dt$.  (Thus, steps 2 and 4 together update the velocities by a full time step to second-order accuracy in a symmetrical way, using the average of the old and new accelerations.)
\end{enumerate}

The simulation will run faster with a larger value of $dt$, but this also reduces the accuracy and, worse, can lead to a runaway instability (exponentially increasing total energy) if $dt$ is too large.  In practice, $dt\sim 0.01$ (in natural units) usually produces good results under the conditions that are of most interest.  

The system's total energy can be calculated at any time, using Eq.~(\ref{LJ}) to calculate the potential energy.  The total energy should be conserved, so monitoring the energy is a good way to check the accuracy of the simulation.

If the system is in thermal equilibrium, then its temperature $T$ should be related to the total kinetic energy $K$ by the equipartition theorem:
\begin{equation}
\label{temperature}
\langle K \rangle = \frac{d}2 Nk_B T,
\end{equation}
where $\langle\,\rangle$ denotes an average and $d$ is the dimensionality of space.  In two dimensions, and in units where $k_B=1$, the temperature is just the average kinetic energy per particle.  Importantly, however, this interpretation is accurate only for systems in thermal equilibrium.  As the system evolves from a nonequilibrium state toward an equilibrium state the ``temperature'' (computed from Eq.~(\ref{temperature})) will drift.  Even after equilibrium is reached, for the relatively small systems ($N\lesssim 1000$) considered here, the instantaneous ``temperature'' will fluctuate significantly and so a time average is needed to obtain an accurate temperature measurement.

With fixed-wall boundary conditions, the instantaneous pressure is simply the total force per unit area (or per unit length in two dimensions) exerted on (or by) the walls.  This too will fluctuate significantly, so a time average is necessary to get a good pressure measurement.

Molecular dynamics simulations are computationally intensive, so programmers must pay attention to calculation times and optimization opportunities.  Traditionally these simulations are coded in Fortran or C, but today it is also feasible to use interpreted languages, such as Java or JavaScript, so long as the interpreter employs just-in-time compilation with good optimization.\cite{browsers, python}  On a personal computer it is now fairly easy to simulate and draw 1000 Lennard-Jones particles at aesthetically pleasing animation rates, and even the current generation of higher-end smartphones and tablets can feasibly simulate and animate hundreds of particles.

The computational bottleneck is in step 3 of the algorithm, which calculates the Lennard-Jones force between each interacting pair of particles.  The number of pairs is $N(N-1)/2$, so the calculation time (per simulated time step) grows in proportion to $N^2$ when $N$ is large.  Programmers using C-derived languages should compute the powers of $r$ in the Lennard-Jones force using repeated multiplication, rather than the much slower ``pow'' function.  Another easy optimization is to truncate the Lennard-Jones interaction at a cutoff of $r\approx3.0$ (in natural units), setting both the force and the potential energy to zero beyond the cutoff (and adding a small constant to the energy within the cutoff, to eliminate the resulting discontinuity); then there is no need to calculate the force for pairs that are separated by more than the cutoff.  The use of a cutoff also makes it possible\cite{CellOptimization} to organize the force calculations in such a way that the calculation time grows only in proportion to~$N$ rather than~$N^2$.

\section{Equilibrium and nonequilibrium states}

Any Lennard-Jones simulation, if it includes graphical output and a way to generate an assortment of initial states, can quickly demonstrate a wide variety of interesting behaviors.

\begin{figure}[b]
\centering
\includegraphics[width=7.5cm]{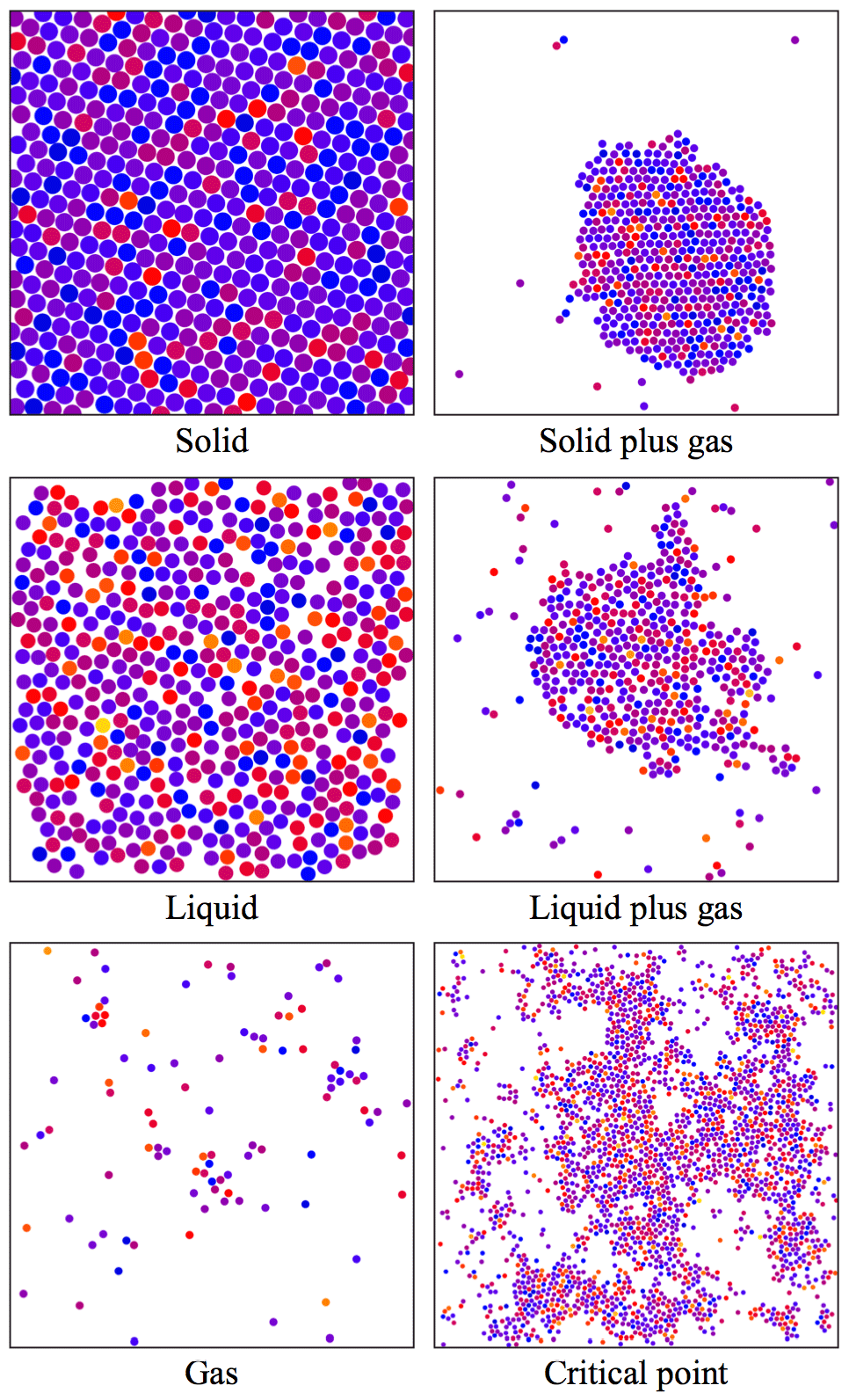}
\caption{Equilibrium states of a two-dimensional Lennard-Jones system.  In this and subsequent images, the particles are drawn as circles of diameter~$\sigma$, colored according to speed with lighter colors corresponding to higher speeds.}
\label{EquilibriumPictures}
\end{figure}

Figure \ref{EquilibriumPictures} shows snapshots of several equilibrium states of the two-dimensional Lennard-Jones system.  In each case the system was simply left to evolve (with constant energy, volume, and number of particles) until its macroscopic properties appeared to be stable.  The gas, liquid, and solid phases are readily apparent at suitable densities and temperatures.  The gas phase can be far from ideal, with small clumps of atoms constantly forming and breaking apart.  A hexagonal crystal structure spontaneously forms to give the rigid solid phase, but forms only partially, over short distances, in the non-rigid liquid phase. Under many conditions this fixed-volume system will settle into an inhomogeneous mixture, with a condensed crystal or droplet constantly exchanging atoms with a surrounding gas.  At one special density (about 0.35 particles per unit volume) and temperature (about 0.50), density variations occur on all possible length scales; this is the critical point.

\begin{figure}[b]
\centering
\includegraphics[width=7.5cm]{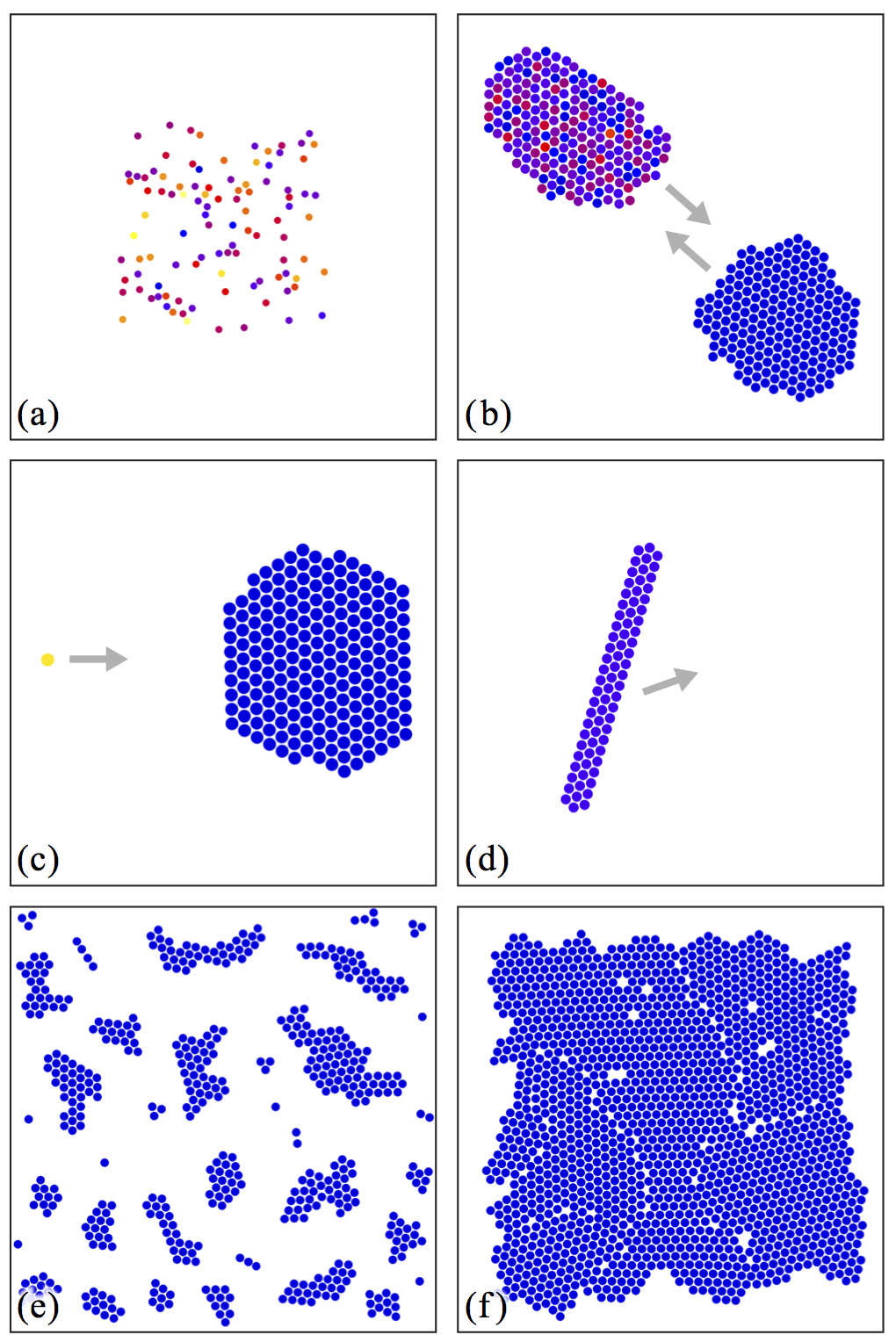}
\caption{Some nonequilibrium states of a two-dimensional Lennard-Jones system. (a) A gas that fills only the central portion of the region, about to expand freely into the surrounding vacuum; (b) hot and cold solids drifting toward one another, which will soon merge and allow heat to flow from one to the other; (c) a fast-moving atom heading toward a frozen crystal, where it will stick and set the crystal into vibrational motion; (d) an elongated crystal drifting uniformly, which will collide with the container walls and begin rotating and vibrating, eventually settling into chaotic Brownian motion; (e) a collection of small crystals, obtained by cooling a gas quickly, which will drift into each other and warm as potential energy is converted to kinetic energy; and (f) a solid that has cooled and crystallized quickly, forming metastable dislocations and domain boundaries.}
\label{NonequilibriumPictures}
\end{figure}

Fascinating as these equilibrium states are, however, they are just the beginning.  With suitable initialization, a Lennard-Jones simulation can also show a huge variety of nonequilibrium states and their subsequent evolution toward equilibrium.  Figure~\ref{NonequilibriumPictures} shows just a few of the possibilities.

\begin{figure}[h]
\centering
\includegraphics[width=7.5cm]{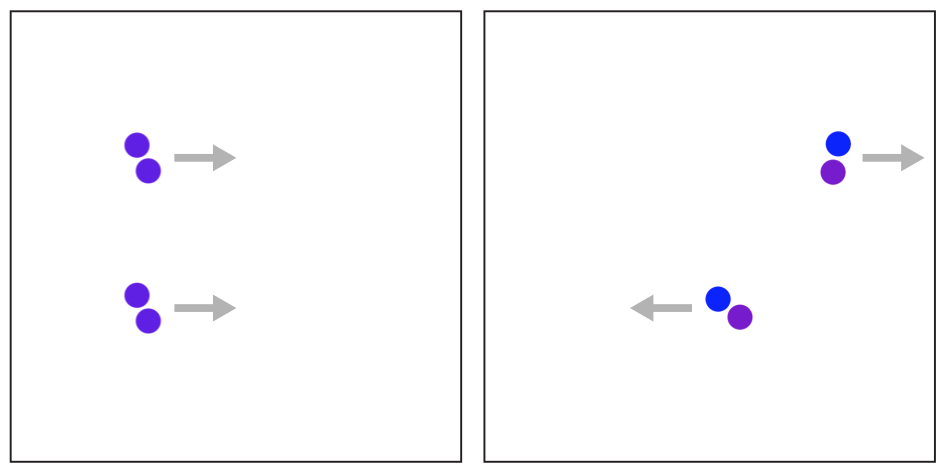}
\caption{A simple demonstration of chaos.  Left: Two diatomic Lennard-Jones ``molecules'' start in identical states of pure horizontal motion, separated vertically so they don't interact with each other.  The molecules subsequently bounce back and forth off the vertical walls, with each bounce converting energy between translational, rotational, and vibrational forms.  Right: Although the motions of the two molecules appear to be perfectly synchronized at first, after several bounces they diverge greatly.  This divergence is caused by the exponential growth of tiny numerical round-off errors.}
\label{ChaoticBouncers}
\end{figure}

Exploring the evolution of nonequilibrium states can give students a vivid understanding of the arrow of time.  Naturally, none of the configurations in Fig.~\ref{NonequilibriumPictures} will ever recur spontaneously after the system has equilibrated.\cite{annealing}  An especially instructive exercise is to reverse all the atoms' velocities after the system has evolved toward equilibrium for a short time, and observe whether the original nonequilibrium state is restored.  Tiny numerical round-off errors, amplified by the chaotic behavior of the system, will almost always set a time limit beyond which the behavior cannot be reversed.  Chaotic behavior occurs even when there are as few as two atoms, as is easily demonstrated using simple configurations such as that shown in Fig.~\ref{ChaoticBouncers}.

\section{User interface}

In order to explore configurations like those shown in Figs.\ \ref{EquilibriumPictures} through \ref{ChaoticBouncers}, students must have ways to put the Lennard-Jones system into a wide variety of initial states.  The easier this process is, the more different behaviors they can discover.  Directly manipulating the state of the system can also make the simulation more interactive and enjoyable.  Of course, there can be good pedagogical reasons to restrict the range of possible configurations, in order to focus students' attention on certain phenomena.  For a simulation to come across as more than a mere movie, however, students should be able to carry out at least some of the following actions:
\begin{itemize}
\item Add and remove atoms.
\item Change the volume of the container.
\item Change the system's energy, e.g., by multiplying all the velocities by a given factor.
\item Drag an individual atom around and/or pull it with a simulated force, using mouse/pointer/touch interactions.
\item Reverse the motions of all the atoms.
\item Exert a uniform ``gravity'' force that pulls all the atoms downward.
\item Save and restore the state of the system.
\item Edit the detailed numerical position and velocity data, or import data from a spreadsheet.
\end{itemize}

There are many possible ways to implement each of these features.  For example, the code that accompanies this article\cite{theSimulation} uses a slider to control the number of atoms, inserting each new atom at the first available open space in a grid; LJfluidApp\cite{OSPApplet} instead requires the user to select the number of atoms from a menu before initialization; Atoms in Motion\cite{AtomsInMotion} uses buttons to add atoms at random locations one at a time; Virtual Molecular Dynamics Laboratory\cite{VMDL} uses mouse clicks to place added atoms; and States of Matter\cite{phet} shows a simulated bicycle pump that sprays new atoms into the container.  Each of these design choices has its advantages and disadvantages, and these choices can have significant effects on the ways that the simulation will be used.

Since the early days of graphical user interfaces, however, experts have recognized qualities of good interface design that help users feel comfortable with software.  Among these qualities are responsiveness, permissiveness, and consistency.\cite{InsideMacintosh}  Responsiveness means that user inputs produce immediate, rather than delayed, effects, so the user can quickly learn what the controls do.  Permissiveness means that the software lets the user perform any reasonable action at any time, rather than imposing arbitrary restrictions,\cite{avoidingInstability} so the user feels in control of the software rather than the other way around.  And consistency means that user-interface controls are easily recognizable, so the user doesn't need to guess at meanings or read lengthy instructions.  While some physicists may consider these qualities to be irrelevant to the content of a simulation, for a student they can make the difference between doing the bare minimum to complete an assignment, and actively engaging to explore widely and learn deeply.

If students are to perform quantitative numerical experiments, the user interface must also include data output.  This output can be as simple as pressure and temperature readouts, or as detailed as a complete dump of all the atoms' position and velocity values.  The exercises in the following section assume that the software provides a means of collecting the required data.\cite{BuiltInPlotting}

\section{Student exercises}

The following exercises are designed for use in undergraduate courses at the beginning and intermediate levels.  They illustrate and reinforce the concepts of thermal physics using numerical data for a nontrivial system, and highlight the idealizations ordinarily made in undergraduate courses, often showing when and how these idealizations break down.  Some of the exercises could be developed further into advanced projects.

All of these exercises can be carried out using the software that accompanies this article.\cite{theSimulation}  Some of them can also be done using other existing software, and further software options will undoubtedly become available in the future.  In most cases, instructors should supplement the exercises with software-specific instructions for configuring the system and acquiring the needed data.\cite{EnhancedExercises}  The exercises  use natural units, but could be augmented by asking students to convert results to conventional units (see Table~I), or adapted for use with software that uses conventional units.  Many of the exercises require that students use a spreadsheet program or other plotting software.

\begin{enumerate}

\item \textbf{Basic phase behavior.}  Run the simulation under a variety of conditions, adjusting the number of atoms, volume, and energy, and waiting for the system to equilibrate at each setting, until you have produced each of the following states of matter:  (a) a pure gas, with plenty of space around most atoms and no large clumps; (b) a pure liquid, with little space between atoms but no long-range order; (c) a pure solid, with all the atoms in orderly rows; (d) a liquid droplet surrounded by gas; (e) a solid crystal surrounded by gas.  For each of these states, write down the following data:  number of particles, volume (actually area in two dimensions), total energy, kinetic energy, potential energy, temperature, and pressure.  Summarize your results in a paragraph, describing the appearance of each phase and any other interesting behavior that you notice.

\item \textbf{Comparison to an ideal gas.}  Set the number of particles ($N$) to 100 and the volume ($V$, actually an area in two dimensions) to approximately 5000.  Add or remove energy until the temperature ($T$) remains stable at approximately 1.0, then note the pressure ($P$) and compute the ratio $PV/Nk_BT$ (remembering that in the natural units used by the simulation, $k_B=1$).  What is this ratio for an ideal gas, and how does your result compare?  Repeat (using the same $N$ and $V$) for $T\approx0.5$ and $T\approx0.3$, being sure to remove enough energy for the system to equilibrate at these approximate temperatures.  Describe the way in which this system's behavior differs from that of an ideal gas, and explain the reason for this difference, noting the visual appearance of the system at each temperature.

\item \textbf{Free expansion.}  Set up an experiment in which a nearly ideal gas (perhaps 100 atoms with an initial volume of about 5000) expands into a vacuum to approximately double its volume.  If the initial temperature is about 2.0, what is the final temperature, and why?  Repeat the experiment with a smaller initial volume (perhaps 1000), and explain the results.  Then try it with a smaller initial temperature as well (perhaps 1.0), and again explain the results.

\item \textbf{Heat capacities and equipartition.}  Use the simulation to measure the heat capacity (at constant volume) of the Lennard-Jones system when it is (a) a nearly ideal gas, and (b) a single solid crystal, with at least 100 atoms in each case.  For each of these measurements you will need to measure the temperatures at \textit{two} slightly different energies ($E$), then subtract these nearby values to obtain $\Delta E$ and $\Delta T$.  Compare your results to the predictions of the equipartition theorem, thinking carefully about how many degrees of freedom the system has.  (Don't expect perfect agreement, but don't expect enormous disagreements either.)

\item \textbf{Heat capacity at constant pressure.}  Use the simulation to measure the heat capacity at constant pressure of the Lennard-Jones system when it is (a) a nearly ideal gas, and (b) a single solid crystal.  In each case you'll have to increase the volume by a small percentage, then add energy until the pressure reaches its previous value.  In calculating the results, be sure to include the $PdV$ term in $C_P = (dE+PdV)/dT$.  For the gas, compare your result to the exact formula derived in textbooks.  For the solid, check that $C_P>C_V$.

\item \textbf{Pressure and energy as functions of temperature.}  For a simulated system of 100 atoms or more, with the volume fixed in the range of 10 to 20 units per atom, measure the total energy, temperature, and pressure over the full range of temperatures from 0 to 1.0, in intervals of 0.1 or less.  Be sure to let the system equilibrate at each temperature before recording your data.  Then plot a graph of pressure vs.\ temperature and another graph of energy vs.\ temperature.  (See Fig.~\ref{PandEvsT} for an example solution.) Comment on the portions of these graphs that can be understood in terms of the ideal gas law and the equipartition theorem, and on the portions that cannot be so simply understood (and why).  How does the low-temperature behavior of the heat capacity differ from that of a real-world solid?

\begin{figure}[t]
\centering
\includegraphics[width=7.5cm]{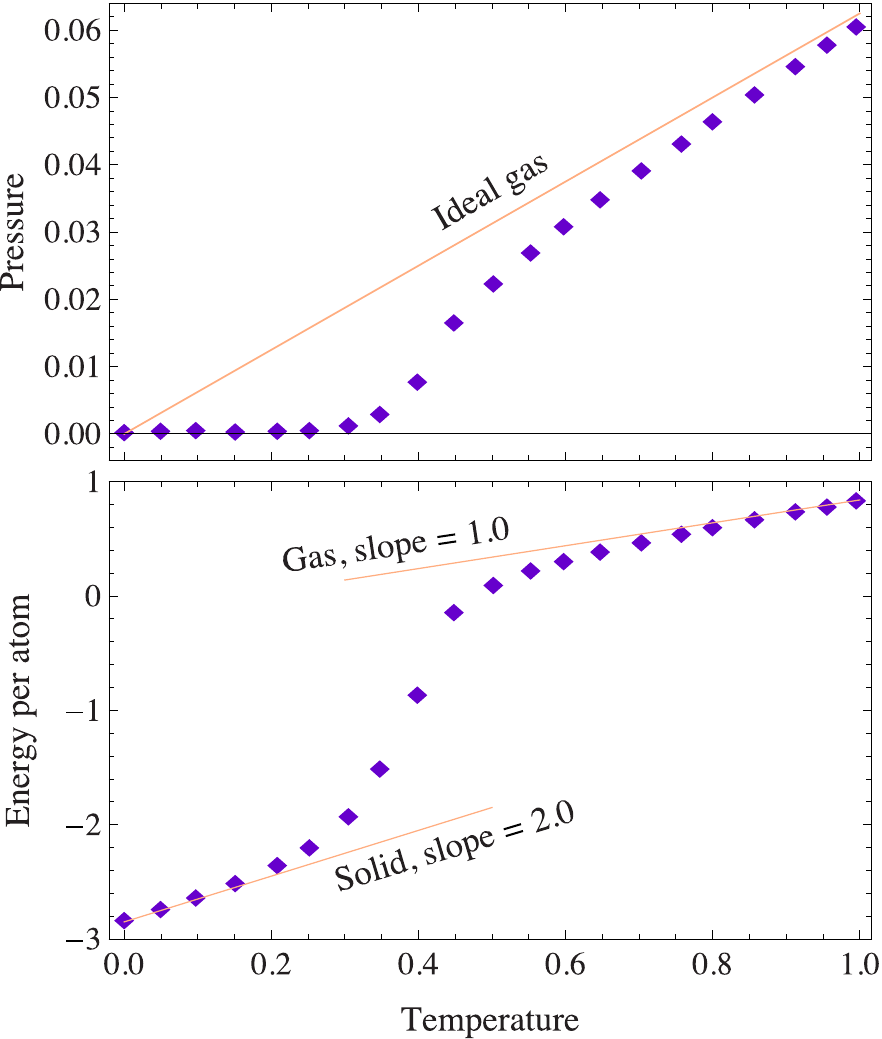}
\caption{Pressure (top) and energy per particle (bottom) as functions of temperature for a simulated system of 100 Lennard-Jones particles in a two-dimensional volume of 1600. All quantities are in natural units. The straight lines show the ideal gas pressure $Nk_BT/V$ and the equipartition predictions for the heat capacities $C$ of a solid ($C/N = 4\cdot\frac12 k_B$) and an ideal gas ($C/N = 2\cdot\frac12 k_B$). See Exercise~6.}
\label{PandEvsT}
\end{figure}

\item \textbf{Heat capacity and entropy.}  From your data in the previous problem, construct a graph of the heat capacity at constant volume, $C_V$, as a function of temperature.  You may have to do some smoothing to reduce the effects of noise in the data.  Then construct a table and graph of $C_V/T$ vs.~$T$, and numerically integrate this function (to an accuracy of one or two significant figures) to determine the entropy as a function of temperature, relative to the entropy at $T=0.1$.  Why can't you determine the \textit{absolute} entropy, relative to $T=0$?  Why doesn't this limitation affect real-world materials?

\item \textbf{Critical point.}  Set the number of atoms to at least 1000 (more is better) and the volume, in natural units, to approximately three times the number of atoms.  Add and remove energy to carefully explore the behavior of the system over the temperature range from about 0.4 to 0.7, and describe how its appearance changes over this range.  What is your best estimate of the critical temperature of this system, and what is the corresponding pressure?

\item \textbf{Phase diagram.}  Map out the approximate phase diagram of the two-dimensional Lennard-Jones system, by adjusting both the temperature and the volume to find the various phase boundary lines (where two phases coexist in equilibrium).  Keep the number of atoms fixed (preferably at 500 or more).  It's easiest to start at a large volume and low temperature, so the system consists of a single solid crystal surrounded by a low-density gas.  Add energy gradually, letting the system equilibrate at various temperatures and noting the temperature and pressure after each equilibration.  Be sure to note the approximate triple point, where the solid crystal (with atoms in orderly rows) melts into a liquid (with no long-range order).  The critical point is the subject of the previous problem.  Finally, reduce the volume (and the energy) to try to locate the solid-liquid phase boundary at pressures somewhat above that of the triple point.  Plot all of your pressure-temperature measurements, sketching in the approximate phase boundary lines and annotating the plot with descriptions of the system's appearance under the various conditions.

\item \textbf{Phase boundary ambiguities.}  Phase boundary lines are sharp (because the properties of the system across a boundary are discontinuous) only in the limit of an infinitely large system.  As a follow-up to the previous problem, explore how the phase boundary locations depend on the volume of the system and on the number of atoms. For example, try plotting the liquid-gas phase boundary for systems with different sizes but similar average densities. Explain the results qualitatively, by considering what fraction of the atoms in the liquid droplet are near the surface.

\item \textbf{Velocity distribution.}  Record the instantaneous velocities ($x$ and $y$ components) for 1000 or more atoms in equilibrium at a temperature of about 0.5 in natural units.  Using a spreadsheet or other software, plot a histogram of the $v_x$ values, using about 20 bins to cover the velocity interval $-2.0$ to $+2.0$.  Do the same for the $v_y$ values.  Also plot the expected results according to the Maxwell-Boltzmann velocity distribution, which for either component $v_i$ is $\sqrt{m/(2\pi k_B T)} \exp({-mv_i^2/2k_B T})$.  (This is the function that, when multiplied by any small velocity interval $dv_i$, gives the probability of finding a single atom within this interval.  To compare it to your simulation results, you'll have to take into account the number of atoms and the sizes of the histogram bins.)  Repeat this whole procedure for a different temperature (adjusting the histogram range if necessary), and discuss the results.  Does it matter whether the simulated material is in a solid, liquid, or gas state?

\item \textbf{Speed distribution.}  As in the previous problem, record the instantaneous velocities of 1000 or more atoms in equilibrium.  Use a spreadsheet to calculate the speed of each atom, plot a histogram of the speeds, and compare to the theoretical prediction (i.e., the two-dimensional Maxwell speed distribution).  (See Fig.~\ref{speedDist} for an example solution.)  Why does the speed distribution equal zero at $v = 0$, where the distributions for $v_x$ and $v_y$ have their peaks?

\begin{figure}[t]
\centering
\includegraphics[width=7.5cm]{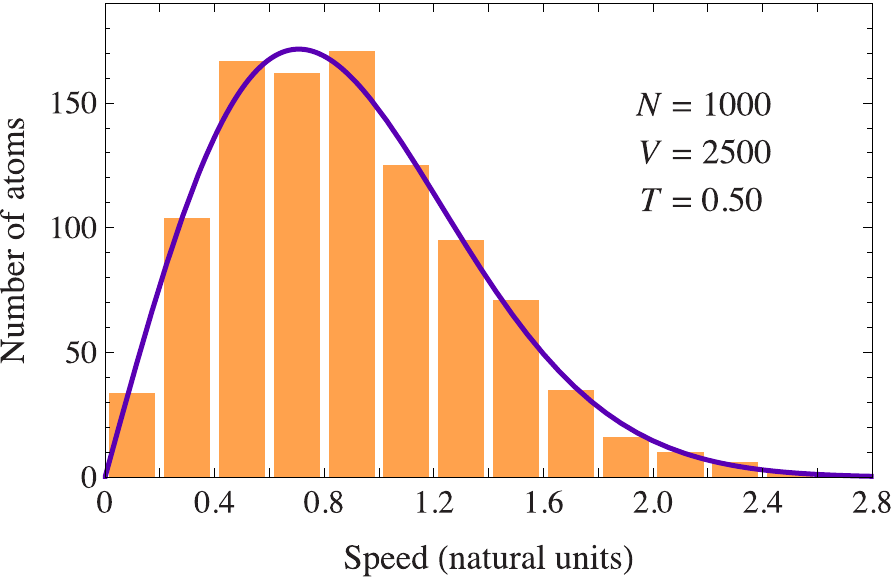}
\caption{Histogram (shown as columns) of the instantaneous speeds of 1000 simulated atoms with $T=0.50$, compared to the prediction of the Maxwell distribution in two dimensions. See Exercise~12.}
\label{speedDist}
\end{figure}

\item \textbf{Gas density in a gravitational field.}  Set the size of the container to $100\times100$, the number of atoms to 500, the gravitational constant to 0.02, and the temperature to about 1.0.  The system is now an ``atmosphere'' whose density decreases with altitude.  How does the typical gravitational energy compare to the typical kinetic energy?  Record the positions of all the atoms at one instant, and use a spreadsheet (or other software) to plot a graph of relative density ($\rho$) as a function of altitude, dividing the full height of the container into ten bins.  (For better statistics you may want to combine data from several ``snapshots'' of the system.)  Fit an exponential function to this data (perhaps plotting it on a semi-log scale), and compare the result to the standard formula for an isothermal atmosphere, $\rho\propto \exp({-mgy/k_BT})$.  \textit{Is} this system isothermal?  Would you \textit{expect} it to be?  Why or why not?  What happens if you reduce the temperature to 0.5?

\item \textbf{Thermal expansion.}  Devise a numerical experiment to measure the thermal expansion of a two-dimensional Lennard-Jones solid.  Explore the temperature range from 0 up to about 0.10, being sure to acquire enough data to see the signal through the statistical noise.  Does the expansion depend linearly on temperature over this range?  What is the approximate thermal expansion coefficient?  How does this behavior compare to that of real solids at low temperatures?

\item \textbf{Brownian motion.}  Set up a simulation of approximately 50 atoms arranged in a stable crystalline shape (not necessarily symmetrical), surrounded by plenty of empty space so the overall volume per atom is 10 or so.  Add or remove energy until the temperature is between 0.06 and 0.08, and run the simulation for a while.  You should see the crystal bounce around randomly, with each bounce off of a wall changing its overall velocity as energy is exchanged between its macroscopic and microscopic degrees of freedom.  Then measure the system's total momentum ($x$ and $y$ components) repeatedly, at regular intervals that are long enough for at least one or two bounces, on average, to occur between measurements. Make at least 100 such measurements (more is better).  From this data set calculate the average values of $p_x^2$ and $p_y^2$, and compare them to the prediction of the equipartition theorem.  Also plot histograms of the $p_x$ and $p_y$ values, and compare them to the prediction of the Maxwell-Boltzmann distribution.

\item \textbf{Brownian bouncing ball.}  As in the previous problem, set up a simulation of a small solid crystal surrounded by empty space.  This time, to minimize the effects of the crystal's orientation, it's best to make it as nearly round as possible.  Freeze the crystal's overall motion when it is near the middle of the region, then turn on a downward gravitational force of 0.001 in natural units.  Run the simulation to let the crystal bounce off the bottom surface a few times, then adjust the temperature to somewhere in the range between 0.06 and 0.10.  Measure the system's total gravitational energy at regular intervals (far enough apart in time for the crystal to move a significant distance).  After making at least a few hundred such measurements, plot a histogram of the gravitational energy values and compare to the prediction of the Boltzmann distribution.

\item \textbf{Fluctuations.}  The simulation calculates temperature by taking a time average of the average kinetic energy per particle. The time average is needed because the \textit{instantaneous} average kinetic energy per particle fluctuates significantly for such a small system.  To study these fluctuations, start with about 50 atoms in a volume of about 250, at a temperature of about 0.4, so the system consists of a liquid droplet surrounded by gas. While the simulation runs, measure the average kinetic energy per particle about a hundred times, and calculate the standard deviation of these measurements. Then repeat this process for systems of about 100, 200, and 500 particles, keeping the density and temperature approximately fixed. How does the standard deviation vary with~$N$? Next, hold $N$ fixed and repeat the process at lower and higher temperatures where the system is entirely solid or entirely gaseous.  Describe and interpret your results as completely as you can.

\item \textbf{Reversibility and chaos.}  Set up configurations similar to those shown in Fig.~\ref{NonequilibriumPictures}(a), (b), and (c), and watch how each evolves over time.  Reverse the motion after a short time, and check whether the reversed motion restores the initial configuration (at least approximately).  In each case, determine the approximate time limit beyond which a reversal will not restore the initial state.  Then set up the configuration of Fig.~\ref{ChaoticBouncers}, and determine the approximate number of bounces before the motions of the two molecules are no longer (approximately) synchronized.

\item \textbf{Thermal conductivity.}  Fill the simulated space with a solid of several hundred atoms, with the left half at a temperature of about 0.1 and the right half at $T=0$.  Starting with this out-of-equilibrium state, step the system forward by small time increments and watch it equilibrate.  Save the state periodically during this process, and use a spreadsheet to calculate the average temperature separately for the left and right halves of the system at each time.  Plot these temperatures vs.\  time, and determine the approximate initial value of the slope of each graph, $dT/dt$.  Finally, use this result to estimate the thermal conductivity, $k_t$, of the two-dimensional Lennard-Jones solid.  This quantity is defined by the two-dimensional version of the Fourier heat conduction law, $Q/\Delta t = -k_t L \,dT/dx$, where $x$ is the coordinate along which the temperature varies, $L$ is the cross-sectional length of the material (measured perpendicular to~$x$), and $Q$ is the amount of heat that flows across the boundary in time $\Delta t$.  You'll need to know the heat capacity of the material, which you can determine from your data or from exercise 4 or 6 above.

\item \textbf{Diffusion.}  Configure the simulation to model a fluid of 1000 atoms, in a volume of 1600, at a temperature of approximately 1.0.  Select one atom that is initially near the center of the container and as the simulation runs, record its position at regular intervals of one time unit, for at least 200 time units.  (If it reaches the edge of the container in less than 200 time units, discard the data and try again.)  Then, using a spreadsheet or other computing environment, compute the squared displacement, $(\Delta x)^2+(\Delta y)^2$, for each of the (200 or so) one-unit time intervals in your data set.  Average these values to obtain the mean squared displacement (MSD).  Similarly, use the same data set to calculate the MSD for time intervals ($\Delta t$) of 2, 5, 10, and 20 units.  Plot the MSD vs.\ $\Delta t$ and notice that the graph is approximately linear; this is the characteristic behavior of diffusive motion (or a so-called random walk).  The slope of the line is closely related to the diffusion constant, $D$; in two dimensions, the MSD is $4D\Delta t$.  Estimate the diffusion constant from your data, then repeat the analysis, holding the fluid density fixed, at temperatures of approximately 0.5 and~2.0.

\end{enumerate}

\section{Enhancements}

As the preceding sections illustrate, the pure Lennard-Jones system exhibits a rich variety of physical behaviors that can keep students occupied almost indefinitely. Still, there are sometimes good reasons to go beyond the pure Lennard-Jones system.

Atoms in Motion,\cite{AtomsInMotion} for example, can simulate arbitrary mixtures of Lennard-Jones particles of five different types, with sizes, masses, and interaction strengths chosen to model helium, neon, argon, krypton, and xenon.  States of Matter\cite{phet} cannot simulate mixtures, but can separately simulate four different types of molecules: two different noble gases, a rigid diatomic species (``oxygen''), and a rigid triatomic species (``water'').  Molecular Workbench\cite{mw} includes an option for modeling charged ions that exert long-range Coulomb forces.

\begin{figure}[b]
\centering
\includegraphics[width=7.5cm]{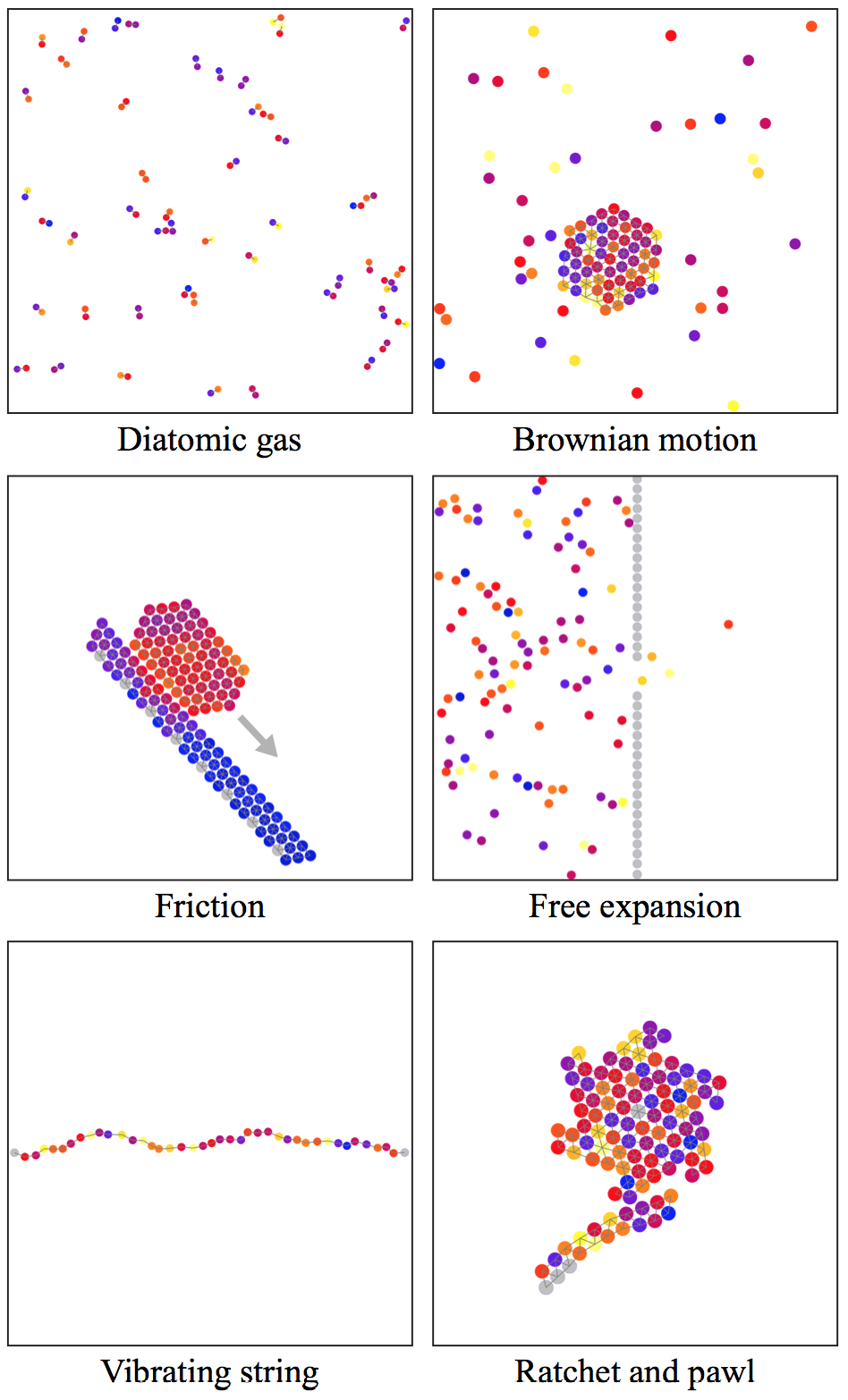}
\caption{A few of the configurations that are possible with a molecular dynamics simulation that allows connecting atoms together with ``bonds'' and anchoring atoms at fixed locations.}
\label{EnhancedPictures}
\end{figure}

In the spirit of encouraging interactive exploration, the software accompanying this article\cite{theSimulation} allows the user to connect any two atoms together with an elastic ``bond'' that adds a spring-like force to the Lennard-Jones force.  This feature is quite versatile and was easy to code.  It is not an accurate model of actual covalent bonds, because there is no limit on the number of bonds per atom, there are no constraints on the angles between bonds, and the bond stiffness, for computational reasons, is unrealistically low.  Still, even this crude model of bonds enables some interesting demonstrations and experiments, such as measuring the heat capacity of a diatomic gas (including the contribution of vibrational potential energy), watching the Brownian motion of a large object bombarded by fast-moving atoms, and observing the friction of one solid object sliding over another (see Fig.~\ref{EnhancedPictures}).

The same simulation also incorporates a second \textit{ad hoc} feature: the ability to anchor an atom so that it is fixed in space.  In this way the user can build barriers and even simulate nano-scale ``machinery'' such as a version of the famous Brownian ratchet.\cite{ratchet}  If nothing else, such demonstrations vividly illustrate how the nano-scale world, with its van der Waals forces and perpetual jiggling motions, differs from the macroscopic world we are used to.

\section{Discussion}

In summary, an interactive molecular dynamics simulation can augment the teaching of thermal physics and related topics in a variety of ways, complementing the more traditional approaches and highlighting some of the idealizations that those approaches require.

On the other hand, any computer simulation incorporates its own set of idealizations and limitations. The simulations described in this article are limited to a rather small number of particles (no more than a few thousand), living in a two-dimensional world.  These simulations are reasonably accurate at modeling only noble gas atoms, and make no attempt to model chemical reactions.

A critical yet intrinsic limitation is that these simulations do not incorporate any quantum effects.  This limitation means that their low-temperature behavior is never realistic, because quantum effects are responsible for the ``freezing out'' of degrees of freedom and other phenomena related to the third law of thermodynamics.  Other approaches\cite{QuantumStatMech} can be used to introduce students to thermodynamic systems at low temperature, at least when the systems are in equilibrium.

No ``canned'' simulation can offer students the same opportunities for open-ended exploration as writing their own code.  It is my hope that, after a certain amount of time spent with the interactive simulations described here---and reaching the limits of what their graphical user interfaces allow---students will be motivated to take the next step and begin modifying the code, or writing their own, to conduct further explorations.

Finally, we should remember that no simulation or numerical ``experiment'' is a substitute for carrying out real experiments on real physical systems.  Rather, a simulation can help bridge the gap between theory and experiment, and often, for thermodynamic systems, between the microscopic and the macroscopic.

\bigskip

\begin{acknowledgments}
I am grateful to Adam Johnston, John Mallinckrodt, Tom Moore, and Paul Weber for their assistance and suggestions regarding various aspects of this article.  This work was supported in many ways by Weber State University.
\end{acknowledgments}


\begin{thebibliography}{99}

\bibitem{FeynmanLectures} Richard P. Feynman, Robert B. Leighton, and Matthew Sands, \textit{The Feynman Lectures on Physics}, volume I (Addison-Wesley, Reading, MA, 1963), available online at \url{<http://www.feynmanlectures.caltech.edu/>}. The ``atomic hypothesis'' quote is on p.~1-2.

\bibitem{FeynmanVideo} For a video interview of Feynman applying his imagination and thinking to the atomic hypothesis, see ``Fun to Imagine I: Jiggling Atoms,'' BBC program first broadcast July 8, 1983, \url{<http://www.bbc.co.uk/archive/feynman/10700.shtml>}, also available at \url{<https://www.youtube.com/watch?v=v3pYRn5j7oI>}.

\bibitem{VMDL} Boston University Center for Polymer Studies, ``Virtual Molecular Dynamics Laboratory,'' \url{<http://polymer.bu.edu/vmdl/>}.  This Web site also introduces molecular dynamics with the Feynman ``atomic hypothesis'' quote, which is too good not to steal.

\bibitem{NovickNussbaum} Shimshon Novick and Joseph Nussbaum, ``Pupils' Understanding of the Particulate Nature of Matter:  A Cross-Age Study,'' Sci. Ed. {\bf 65} (2), 187--196 (1981).

\bibitem{MakingSense} Rosalind Driver \textit{et al.}, \textit{Making Sense of Secondary Science: Research into Children's Ideas} (Routledge, London, 1994), Chapter 11.

\bibitem{Arons} Arnold B. Arons, \textit{A Guide to Introductory Physics Teaching} (Wiley, New York, 1990), pp.\ 274--281.

\bibitem{AllenAndTildesley} M. P. Allen and D. J. Tildesley, \textit{Computer Simulation of Liquids} (Clarendon Press, Oxford, 1987).

\bibitem{Haile} J. M. Haile, \textit{Molecular Dynamics Simulation} (Wiley, New York, 1992).

\bibitem{Rapaport} D. C. Rapaport, \textit{The Art of Molecular Dynamics Simulation}, second edition (Cambridge University Press, Cambridge, 2004).

\bibitem{AtomicMicroscope} Stark Design, ``Atomic Microscope'' (Windows and Mac Classic application).  This software, first released around 1999, is apparently no longer available, but is described in Ref.~\onlinecite{AtomicMicroscopeReview} and is similar in many ways to Atoms in Motion, Ref.~\onlinecite{AtomsInMotion}.

\bibitem{AtomicMicroscopeReview} Shawn Carlson, ``Modeling the Atomic Universe,''
Sci. Am. {\bf 281} (4), 118--119 (Oct.\ 1999), \url{<http://www.scientificamerican.com/article.cfm?id=modeling-the-atomic-unive>}.

\bibitem{AtomsInMotion} Atoms in Motion LLC, ``Atoms in Motion'' (iPad app), \url{<http://www.atomsinmotion.com/>} (2011--2012).

\bibitem{phet} PhET project, ``States of Matter'' (Java Web Start application), \url{<http://phet.colorado.edu/en/simulation/states-of-matter>} (2009--2012).

\bibitem{mw} The Concord Consortium, ``Molecular Workbench'' (Java simulations and modeling tools), \url{<http://mw.concord.org/>} (2004--2013). A new HTML5 version is also available, at \url{<http://mw.concord.org/nextgen/>}.

\bibitem{GouldTobochnikChristian} Harvey Gould, Jan Tobochnik, and Wolfgang Christian, \textit{An Introduction to Computer Simulation Methods}, third edition (Pearson Addison Wesley, San Francisco, 2007).

\bibitem{Giordano} Nicholas J. Giordano and Hisao Nakanishi, \textit{Computational Physics}, second edition (Pearson Prentice Hall, Upper Saddle River, NJ, 2006).

\bibitem{DVSJavaManual} Daniel V. Schroeder, \textit{Physics Simulations in Java: A Lab Manual} (unpublished, 2006--2011), \url{<http://physics.weber.edu/schroeder/javacourse/>}.

\bibitem{Sander} Leonard M. Sander, \textit{Equilibrium Statistical Physics: With Computer Simulations in Python} (CreateSpace Independent Publishing, 2013), \url{<http://www-personal.umich.edu/~lsander/ESP/ESP.htm>}. 

\bibitem{GouldTobochnikStat} Harvey Gould and Jan Tobochnik, \textit{Statistical and Thermal Physics: With Computer Applications} (Princeton University Press, Princeton, 2010).

\bibitem{OSPApplet} Open Source Physics Project, ``LJfluidApp,'' \url{<http://stp.clarku.edu/simulations/lj/md/index.html>} (2009).

\bibitem{theSimulation} See supplemental material at \url{<http://dx.doi.org/10.1119/1.4901185>} for the simulation program, user instructions, and a version of the exercises in Section~VI that is customized for this particular simulation. These materials are also available at \url{<http://physics.weber.edu/schroeder/md/InteractiveMD.html>}.

\bibitem{london} In this case the van der Waals force is also called the London dispersion force; it is caused by quantum fluctuations of the molecules' electronic charge distributions.  The $1/r^6$ dependence is derived using perturbation theory in many quantum mechanics textbooks. See also Barry R. Holstein, ``The van der Waals interaction,'' Am. J. Phys. \textbf{69} (4), 441--449 (2001); and Kimball A. Milton, ``Resource Letter VWCPF-1: van der Waals and Casimir-Polder forces,'' Am. J. Phys. \textbf{79} (7), 697--711 (2011).

\bibitem{maitland} Geoffrey C. Maitland \textit{et al.}, \textit{Intermolecular Forces: Their Origin and Determination} (Clarendon Press, Oxford, 1981).  This monograph makes an exhaustive assessment of the limitations of the Lennard-Jones 6-12 potential.

\bibitem{WallsComment} The walls can be hard, producing instantaneous reversals in velocity, or soft, exerting a spring-like repulsive force that grows as atoms penetrate into the walls more deeply.  The accompanying code\cite{theSimulation} uses soft walls (with a spring constant of 50 in natural units), because of the simplicity of all forces being smoothly varying functions of position.  A disadvantage of this method is that the volume of the simulated space is somewhat variable and hence ambiguous, especially at high temperatures.

\bibitem{browsers} As of this writing, the current versions of all major browsers for personal computers deliver impressive JavaScript performance that is more than adequate for the simulations described in this article.  Performance of JavaScript on mobile devices, however, is much more variable.

\bibitem{python} Another popular interpreted language is Python, which offers advantages in a computational physics course or other setting where students will be writing or modifying the code. As of this writing, Python's relatively poor performance limits the size of an interactive molecular dynamics simulation running at a reasonable animation rate. However, through use of the NumPy library (\url{<www.numpy.org>}) to vectorize the calculations (see, e.g., the program \texttt{simplemd.py} in Ref.~\onlinecite{Sander}), one can still reach a performance level that is adequate for most of the examples and exercises in this article.

\bibitem{CellOptimization} The most powerful optimization technique is to divide the simulation space into a grid of ``cells'' whose widths are no smaller than the cutoff distance.  Then each atom can interact only with atoms in its own cell and the eight nearest neighbor cells.  The algorithm is described in Refs.~\onlinecite{AllenAndTildesley} and~\onlinecite{Rapaport} and is used in the code of Ref.~\onlinecite{theSimulation}.  The additional coding can be done in only a few dozen lines, and is well worth the trouble for simulations of 500 or more particles, but provides no benefit at all when $N\lesssim100$.

\bibitem{annealing} The configuration shown in Fig.~\ref{NonequilibriumPictures}(f) will not spontaneously equilibrate, but it can be annealed by gradually adding energy.

\bibitem{InsideMacintosh} Apple Computer, Inc., \textit{Inside Macintosh} (Addison-Wesley, Reading, MA, 1985), p.~I-27.

\bibitem{avoidingInstability} Implementing ``permissiveness'' in a molecular dynamics simulation can be challenging, because some potential user actions (e.g., placing the atoms so they overlap) can add large amounts of energy to the system, triggering the numerical instability described in Section~III.  The accompanying simulation\cite{theSimulation} tries to adapt to dangerous user actions by decreasing the time step $dt$ and limiting the rate at which particles can be manually pushed together.  Still, it isn't hard for a curious user to ``break'' the simulation, generating an error message and necessitating a reset.  This experience can sometimes be instructive but is usually just frustrating.

\bibitem{BuiltInPlotting} Some molecular dynamics software\cite{VMDL, OSPApplet} goes further to include built-in plotting of various data.  This feature can be useful for quick demonstrations, but it usually reduces the degree to which the student is actively engaged in deciding how to gather and analyze the data.  Also, as a practical matter, it is difficult to pre-program all of the different types of plots that students and instructors might wish to make.

\bibitem{EnhancedExercises} The online supplement\cite{theSimulation} to this article includes a version of the exercises with instructions and hints that are specific to the accompanying software.

\bibitem{ratchet} Feynman \textit{et al.}, Ref.~\onlinecite{FeynmanLectures}, Chapter 46.  See also Harvey S. Leff and Andrew F. Rex, \textit{Maxwell's Demon 2: Entropy, Classical and Quantum Information, Computing} (Institute of Physics Publishing, Bristol, 2003), Section 1.2.5, and references therein.

\bibitem{QuantumStatMech}  Besides the usual approaches to quantum statistical mechanics found in every textbook, see (for example) Thomas A. Moore and Daniel V. Schroeder, ``A different approach to introducing statistical mechanics,'' Am. J. Phys. {\bf 65} (1), 26--36 (1997); Martin Ligare, ``Numerical analysis of Bose-Einstein condensation in a three-dimensional harmonic oscillator potential,'' Am. J. Phys. {\bf 66} (3), 185--190 (1998); and J. Arnaud \textit{et al.}, ``Illustration of the Fermi-Dirac statistics,'' Am. J. Phys. {\bf 67} (3), 215--221 (1999).

\end{thebibliography}
\end{document}